\documentstyle[12pt]{article}

\newcommand{\be}{\begin{equation}}
\newcommand{\ee}{\end{equation}}
\newcommand{\prt}{\partial}
\newcommand{\vp}{\varphi}
\newcommand{\br}{{\bf r}}
\newcommand{\om}{\omega}
\newcommand{\Dlt}{\Delta}
\newcommand{\al}{\alpha}
\newcommand{\bt}{\beta}
\newcommand{\dlt}{\delta}
\newcommand{\gm}{\gamma}
\newcommand{\ep}{\varepsilon}

\begin{document}

\begin{center}
{\large{\bf Coherent Resonance in Trapped Bose Condensates} \\ [2mm]
V.I. Yukalov$^{1,2}$, E.P. Yukalova$^{1,3}$, and V.S. Bagnato$^1$}\\ [3mm]

{\it $^1$Instituto de Fisica de S\~ao Carlos, Universidade de S\~ao Paulo \\
Caixa Postal 369, S\~ao Carlos, S\~ao Paulo 13560-970, Brazil \\ [3mm]
$^2$Bogolubov Laboratory of Theoretical Physics \\
Joint Institute for Nuclear Research, Dubna 141980, Russia \\ [3mm]
$^3$Department of Computational Physics, Laboratory 
of Information Technologies \\
Joint Institute for Nuclear Research, Dubna 141980, Russia}
\end{center}

\vskip 3cm

\begin{abstract}

Coherent resonance is the effect of resonant excitation of 
nonlinear coherent modes in trapped Bose condensates. This 
novel effect is shown to be feasible for Bose-condensed trapped 
gases. Conditions for realizing this effect are derived. A method 
of stabilizing Bose condensates with attractive interactions is 
advanced. The origin of dynamic critical phenomena is elucidated. 
Interference effects are studied. The existence of atomic squeezing 
and multiparticle coherent entanglement is demonstrated. Coherent 
resonance is a generalization of atomic resonance, involving internal 
states of an individual atom, to collective states of a multiparticle 
system.

\end{abstract}

\newpage

\section{Introduction}

Resonant interaction of an alternating electromagnetic field with 
an atom makes it possible to select a pair of atomic levels and to 
treat the atom as an effectively two-level system [1]. This resonant 
interaction is in the basis of one of the main directions of optics. 
The possibility of realizing atomic resonance is due to the existence 
of discrete energy levels of electrons in an atom, the energy spectrum 
being described by the Schr\"odinger equation. The Bose-condensed 
atoms are also described by the Schr\"odinger equation, though 
nonlinear, which is often termed the Gross-Pitaevskii equation [2--4]. 
Trapped atoms, because of a confining potential, also possess a 
discrete spectrum [5]. A specific feature is that this is a spectrum 
of collective nonlinear states of an ensemble of coherent atoms. The 
collective coherent and nonlinear nature of these states is what makes 
them principally different from single-particle linear states of an 
individual atom. To emphasize this difference, the collective states
of Bose-condensed atomic gases are called {\it nonlinear coherent modes} 
[5]. Since the spectrum of these modes, pertaining to trapped atoms, 
is discrete, it has been suggested [5] that a resonant coupling of two 
such modes is feasible by means of an alternating field with a frequency
tuned to the transition frequency between the chosen levels. This type 
of resonance for exciting nonlinear coherent modes can be named {\it
coherent resonance}. This phenomenon, if realised, could open a whole 
new field of possible applications, analogous to those accomplished with
atomic resonance.

In the present communication, we investigate the conditions required 
for realizing the coherent resonance. We propose a new method for
stabilizing Bose-condensed gas with attractive interactions by
transferring atoms to an excited coherent mode. We study dynamic 
critical phenomena, discovered numerically [6--8], and give their
explanation. We also describe some new properties of Bose condensate
subject to coherent resonance, such as interference effects, atomic 
squeezing, and multiparticle coherent entanglement.

\section{Coherent Resonance}

We consider a system of trapped Bose-condensed atoms at low temperature,
when the Bose gas in a coherent state. The coherent-field wave function
satisfies the Gross-Pitaevskii equation
\be
\label{1}
i\hbar \; \frac{\prt}{\prt t}\; \vp(\br,t) =\left ( \hat H[\vp] +
\hat V\right ) \vp(\br,t) \; ,
\ee
in which the coherent field is normalized to unity, $||\vp||^2=1$. The
nonlinear Hamiltonian is
\be
\label{2}
\hat H[\vp] \equiv -\; \frac{\hbar^2{\bf\nabla}^2}{2m_0} +
U(\br) + NA_s |\vp(\br,t)|^2 \; ,
\ee
where $U(\br)$ is a trapping potential; $m_0$ is atomic mass; $N$ is the
number of atoms; $A_s\equiv 4\pi\hbar^2 a_s/m_0$, with $a_s$ being a
scattering length. The alternating field is
\be
\label{3}
V(\br,t) = V_1(\br)\cos\om t  + V_2(\br)\sin\om t \; .
\ee
Note that Eq. (1) is an {\it exact} equation for the coherent field 
[9], and $\vp(\br,t)$ does not need to be interpreted as an average 
of a field operator, for which case Eq. (1) would be only a mean-field 
approximation.

The {\it nonlinear coherent modes} are the stationary solutions to the
eigenproblem
\be
\label{4}
\hat H[\vp_n]\;\vp_n(\br) = E_n\;\vp_n(\br) \; .
\ee
Since the Hamiltonian (2) is nonlinear, the modes are not necessary 
orthogonal to each other, so that the scalar product
$$
(\vp_m,\vp_n) \equiv \int \vp_m^*(\br) \; \vp_n(\br) \; d\br 
$$
is not compulsory a Kroneker delta. But each mode can always be 
normalized, $||\vp_n||^2=1$. Some properties of the nonlinear modes 
have been discussed [5--8,10--13] and dipole mode has been observed 
experimentally [14].

Let us select two spectrum levels, with the energies $E_1$ and $E_2$,
such that $E_1<E_2$. The related transition frequency is
\be
\label{5}
\om_{21} \equiv \frac{1}{\hbar}\;( E_2 - E_1 ) \; .
\ee
The frequency of the alternating field (3) is tuned close to this
transition frequency (5), which implies the resonance condition
\be
\label{6}
\left | \frac{\Dlt\om}{\om}\right | \ll 1 \; , \qquad
\Dlt\om \equiv \om - \om_{21} \; .
\ee

The solution to the evolution equation (1) can be presented in the form
\be
\label{7}
\vp(\br,t) = \sum_n c_n(t)\; \vp_n(\br) \; \exp\left ( -\;
\frac{i}{\hbar}\; E_n\; t\right ) \; .
\ee
We assume that $c_n(t)$ is slow as compared to the exponential, such
that
\be
\label{8}
\frac{\hbar}{E_n}\; \left | \frac{dc_n}{dt}\right | \ll 1 \; .
\ee

There are two transition amplitudes in the system, one is the {\it
internal transition amplitude}
\be
\label{9} \al_{mn} \equiv A_s \; \frac{N}{\hbar}\;
\int |\vp_m(\br)|^2 \left [ 2|\vp_n(\br)|^2 - |\vp_m(\br)|^2
\right ] \; d\br \; ,
\ee
caused by atomic interactions, and another is the {\it external 
transition amplitude}
\be
\label{10}
\bt_{mn} \equiv \frac{1}{\hbar}\; \int \vp_m^*(\br) \left [ V_1(\br) -
i V_2(\br) \right ] \; \vp_n(\br) \; d\br \; ,
\ee
due to the driving resonant field. Similarly to atomic resonance, in 
order to avoid power broadening, the transition aplitudes are to be 
smaller than the transition frequencies,
\be
\label{11}
\left | \frac{\al_{mn}}{\om_{mn}}\right | \ll 1 \; , \qquad
\left | \frac{\bt_{mn}}{\om_{mn}}\right | \ll 1 \; .
\ee
Under conditions (6), (8), and (11), the Gross-Pitaevskii equation (1) 
reduces to the set of equations
\be
\label{12}
i\; \frac{dc_1}{dt} = \al_{12} |c_2|^2 c_1 + 
\frac{1}{2}\; \bt_{12} c_2 e^{i\Dlt\om t} \; , \qquad
i\; \frac{dc_2}{dt} = \al_{21} |c_1|^2 c_2 + 
\frac{1}{2}\; \bt^*_{12} c_1 e^{-i\Dlt\om t} 
\ee
for an effective two-mode system, where $|c_1|^2+|c_2|^2=1$. This 
reduction is valid provided all transition amplitudes, involved in
Eq. (12), are small in the sense of condition (11), which yields the
inequalities
\be
\label{13}
\left | \frac{\al_{12}}{\om_{21}} \right | \ll 1 \; , \qquad
\left | \frac{\al_{21}}{\om_{21}} \right | \ll 1 \; , \qquad
\left | \frac{\bt_{12}}{\om_{21}} \right | \ll 1 \; .
\ee

To analyse the above inequalities, we consider a cylindrical trap 
modelled by the harmonic trapping potential
$$
U(\br) = \frac{m_0}{2}\left ( \om_r^2\; r_x^2 + \om_r^2 \; r_y^2 +
\om_z^2\; r_z^2 \right ) \; ,
$$
with the aspect ratio
\be
\label{14}
\nu \equiv \om_z/\om_r \; .
\ee
Introduce the dimensionless coupling parameter
\be
\label{15}
g \equiv 4\pi\; \frac{a_s}{l_r}\; N \qquad \left (
l_r \equiv \sqrt{\frac{\hbar}{m_0\om_r}} \right ) \; .
\ee
The spectrum of nonlinear coherent modes is defined by the 
eigenproblem (4). The external transition amplitude (10) can always 
be made sufficiently small by regulating the amplitude of the resonant 
field. We need to calculate the internal transition amplitude (9) and 
to compare it with a chosen transition frequency (5). Calculations can 
be accomplished by means of the optimized perturbation theory [15--17] 
(for reviews see [18,19]). The first two inequalities (13) can be valid 
only outside the radius of convergence of the strong-coupling expansion.
This imposes the restriction
\be
\label{16}
|g\nu| < \frac{[2p^2+(q\nu)^2]^{5/4}}{14p\sqrt{q}\; I_{nmj}}
\ee
on the coupling parameter (15). Here the integral 
$I_{nmj}\sim (|\psi_{nmj}|^2,|\psi_{nmj}|^2)$, and  $p\equiv 2n+|m|+1$, 
$q\equiv 2j+1$ are the combinations of  the radial, $n$, azimuthal, $m$, 
and axial, $j$, quantum numbers. The restriction (16) can be rewritten 
as the limitation on the admissible number of particles $N<N_0$, with 
the limiting number
\be
\label{17}
N_0 = \frac{[2p^2 +(q\nu)^2]^{5/4}}{56\pi\nu p\sqrt{q}\; I_{nmj}}\;
\left | \frac{l_r}{a_s}\right | \; .
\ee
For the ground-state level, this reduces to
\be
\label{18}
N_0 = \sqrt{\frac{\pi}{2}}\; \frac{(2+\nu^2)^{5/4}}{14\nu} \;
\left | \frac{l_r}{a_s}\right | \; .
\ee
Thus, the number of particles, which can be resonantly transferred 
to an excited coherent mode, is limited by a number $N_0$, 
which depends on the quantum numbers of the coupled modes. The number 
$N_0$ also essentially depends on the trap shape through the aspect 
ratio (14).

It turns out that the limiting numbers (17) and (18) are close to the
critical values defining the maximal number of atoms with a negative
scattering length, which can be condensed in a trap. Examples of atoms
with attractive interactions are $^7$Li (see review [20]) and $^{85}$Rb
(see Ref. [21]). Such atoms can form condensates only if their number
does not exceed a critical value [5,22--25]. In that case, the condensate
forms a long-lived metastable state, with a very slow decay, caused by
quantum tunneling [26--28], with the tunneling probability being
negligible as compared to the probability of escaping from the trap 
because of depolarizing collisions [23,29]. But if the number of atoms
is larger than critical, the condensate collapses. With a supply of 
atoms from an external source, say by a continuous loading from another
trap [30], the condensate grows again and, after surpassing the critical
number, again collapses. Thus a series of growths and collapses take 
place, as was observed experimentally [31,32] in Bose-condensed $^7$Li
and described theoretically by means of the Gross-Pitaevskii equation
supplemented by relaxation terms and by the related rate equations.
Explosion of an attractive condensate of $^{85}$Rb atoms by manipulating 
the atomic interactions with an external magnetic field near a Feshbach
resonance [36] has also been observed [21,37].

To increase the number of atoms in a condensate with attractive 
interactions, it was suggested [38] to drive a quadrupole collective 
excitation. We may notice that transferring atoms to an excited nonlinear
coherent mode may also stabilize such a condensate. This follows from 
the form (17), which shows that the limiting number $N_0$ increases for 
higher excited modes as
$$
N_0 \sim (2n +|m| + 1)^{3/2} \; , \qquad N_0 \sim (2j+1)^2 \; .
$$
To estimate the number of atoms for highly excited modes, one could 
also use an optimized quasiclassical approximation [39]. As Eq. (18) 
demonstrates, the limiting number $N_0$ is larger for essentially
anisotropic traps, that is, for cigar-shape $(\nu\ll 1)$ and disk-shape
$(\nu\gg 1)$ traps.

The temporal behaviour of the system under coherent resonance is 
described by the evolution equations (12). Numerical solution of these
equations, for the varying amplitude of the resonant field, displays
{\it dynamical critical phenomena} [6--8]. Here we present a general 
picture of these phenomena and elucidate their origin.

Let us present the population amplitudes in the form
\be
\label{19}
c_1 =\sqrt{\frac{1-s}{2}}\; \exp\left\{ i\left ( \pi_1 +
\frac{\Dlt\om}{2}\; t\right ) \right \} \; , \qquad
c_2 =\sqrt{\frac{1+s}{2}}\; \exp\left\{ i\left ( \pi_2 -\;
\frac{\Dlt\om}{2}\; t\right ) \right \} \; ,
\ee
where $\pi_1=\pi_1(t)$ and $\pi_2=\pi_2(t)$ are the phases. The variable
\be
\label{20}
s\equiv |c_2|^2 - |c_1|^2
\ee
is the population difference. Introduce the notation for the average 
internal transition amplitude
\be
\label{21}
\al \equiv \frac{1}{2}\; ( \al_{12} + \al_{21} ) 
\ee
and let us take into account that the external transition amplitude (10)
is, in general, complex valued,
\be
\label{22}
\bt_{12} = \bt e^{i\gm} \; , \qquad \bt \equiv |\bt_{12}| \; .
\ee
Also, define the effective detuning
\be
\label{23}
\dlt \equiv \Dlt\om + \frac{1}{2}\; (\al_{12} - \al_{21} ) \; .
\ee
And introduce the phase difference
\be
\label{24}
x \equiv \pi_2 - \pi_1 +\gm \; .
\ee
Substituting Eqs. (19) into the evolution equations (12), we come to
the set of equations for the population difference (20) and phase 
difference (24),
\be
\label{25}
\frac{ds}{dt} = -\bt\sqrt{1-s^2}\; \sin x \; , \qquad
\frac{dx}{dt} = \al s + \frac{\bt s}{\sqrt{1-s^2}}\; \cos x + \dlt \; .
\ee

Numerical solution of Eqs. (25) demonstrates the existence of {\it 
dynamical critical phenomena}, when a tiny variation of parameters 
results in sharp changes of temporal behaviour of $s(t)$ and $x(t)$. 
We illustrate this in the following figures, where time is measured 
in units of $\al^{-1}$. There are two independent dimensionless 
parameters
\be
\label{26}
b\equiv \frac{\bt}{\al} \; , \qquad \ep\equiv \frac{\dlt}{\al} \; ,
\ee
which can be varied. In what follows, we vary $b$ in the interval 
$[-1,1]$ and keep $|\ep|\ll 1$. And there are also two initial conditions
$s_0=s(0)$ and $x_0=x(0)$.

In Fig. 1, we set $s_0=-1$, $x_0=0$, and $\ep=0$, while varying the
parameter $b$ corresponding to the amplitude of the pumping resonant 
field. In the range $0<b<0.5$, the population difference oscillates 
between $-1$ and zero and the phase difference monotonically decrease, 
as is shown in Fig. 1a and 1b. The critical point $b_c=0.5$ separates 
the regions of two qualitatively different temporal behaviours. After
$b$ surpasses $b_c$, the population difference starts oscillating 
between $-1$ and $+1$, while the phase difference becomes a periodic 
function, as is demonstrated in Figs. 1c to 1e.

As follows from the definition of the internal transition amplitude 
(9), its value may become negative, either for strongly energetically 
separated modes or for atoms with attractive interactions, when $A_s<0$.
In such a case, the parameter (21) is negative, $\al<0$, and so is the
parameter $b<0$ from Eq. (26). The variation of $b$ in the negative 
region is also accompanied by dynamic critical phenomena, as for positive
$b$, though the behaviour of the phase difference is not the same as for 
$b>0$. This is illustrated by Fig. 2, where we set the same initial 
conditions $s_0=-1$ and $x_0=0$, and zero detuning. For $b$ in the 
interval $-0.5<b<0$, the population difference is in the range 
$-1\leq s\leq 0$, as is shown in Figs. 2a and 2b. The critical point 
now is $b_c=-0.5$. After crossing the critical point, when $b<b_c$, 
the amplitude of the population difference increases by a jump, so 
that $s(t)$ oscillates now in the diapason $-1\leq s\leq 1$. The phase
difference is an oscillating function for both $b>b_c$ and $b<b_c$, but 
its shape changes drastically when crossing $b_c$, as can be seen in 
Figs. 2c and 2d.

In this way, there are two critical lines for each given pair of initial
conditions $s_0$ and $x_0$. Changing initial conditions changes the 
location of the critical line on the manifold of the parameters $b$ 
and $\ep$. Thus, in Fig. 3, we show the course of events for the 
initial conditions $s_0=-0.8$ and $x_0=0$, with the detuning $\ep=0$ 
and for the parameter $b$ varying in the whole range $-1\leq b\leq 1$. 
Under the fixed detuning $\ep=0$ and given initial conditions, there 
are two critical points $b_{c1}=-0.8$ and $b_{c2}=0.2$.

For a nonzero detuning, the location of the critical line changes, but
the overall picture remains the same. A finite detuning also makes the
shape of the function $x(t)$ a little asymmetric, as is shown in Fig. 4.

To understand the origin of these critical phenomena, we studied the
phase portrait on the plane $\{ s,x\}$ for different parameters $b$ 
and $\ep$. Examples are demonstrated in Fig. 5. We also accomplished 
the stability analysis of fixed points. This investigation clarify 
the origin of the dynamic critical phenomena. The latter are related 
to the existence of the {\it saddle separatrix} defined by the equation
\be
\label{27}
\frac{s^2}{2} - b\sqrt{1-s^2}\; \cos x + \ep s = b \; .
\ee
The separatrices divide the phase plane $\{ s,x\}$ into the regions 
with different dynamics. As is evident from Eq. (27), changing 
any of the parameters $b$ or $\ep$ shifts the separatrices. When 
a separatrix crosses an initial point $\{ s_0,x_0\}$, the system 
trajectory passes to another region of the phase plane, as a result of
which the system dynamics changes qualitatively. The separatrix crossing
of an initial point corresponds to the {\it critical line}
\be
\label{28}
\frac{s^2_0}{2} - b\sqrt{1-s^2_0}\; \cos x_0 + \ep s_0 = |b| 
\ee
on the manifold of the parameters $b$ and $\ep$. It is this {\it
separatrix crossing effect} that causes the appearance of the dynamic 
critical phenomena. It can be shown [7,8] that a time-averaged system
displays critical phenomena, typical of phase transitions in stationary
statistical systems, on the same critical line.

There exist several more interesting effects occurring in the process 
of coherent resonance. Not to overload the present communication, we 
mention these effects only briefly. A more detailed consideration will 
be done in separate publications.

\subsection{Interference effects}

Coherent resonance couples two selected nonlinear modes, leaving other 
modes unpopulated. Therefore, the coherent field (7) is, effectively, 
a sum
\be
\label{29}
\vp(\br,t) = \vp_1(\br,t) + \vp_2(\br,t) 
\ee
of two terms
$$
\vp_j(\br,t) = c_j(t) \vp_j(\br) \exp\left ( -\; \frac{i}{\hbar}\;
E_j\; t\right ) \qquad (j=1,2) \; .
$$
Using this, one can define the {\it interference pattern}
\be
\label{30}
\rho_{int}(\br,t) \equiv \rho(\br,t) - \rho_1(\br,t) -
\rho_2(\br,t) \; ,
\ee
in which
$$
\rho(\br,t) = |\vp(\br,t)|^2 \; , \qquad \rho_i(\br,t) = 
|\vp_i(\br,t)|^2 \; ,
$$
and the {\it interference current} 
\be
\label{31}
{\bf j}_{int}(\br,t) \equiv {\bf j}(\br,t) - {\bf j}_1(\br,t) - 
{\bf j}_2(\br,t) \; ,
\ee
where
$$
{\bf j}(\br,t) = \frac{\hbar}{m_0} \; {\rm Im}\; \vp^*(\br,t)
{\bf\nabla}\vp(\br,t) \; , \qquad
{\bf j}_i(\br,t) = \frac{\hbar}{m_0} \; {\rm Im}\; \vp^*_i(\br,t)
{\bf\nabla}\vp_i(\br,t) \; .
$$
These quantities show that the total density of atoms in a trap is 
not simply a sum of the partial mode densities but it includes also 
interference fringes and that, because of the different mode topology, 
there arises a local interference current. The interference effects 
can be observed experimentally.

\subsection{Atomic squeezing}

The evolution equations for the effective two-mode system, obtained by 
means of the coherent resonance, can be interpreted as the equations for
the statistical averages of quasispin operators
$$
S_\al \equiv \sum_{i=1}^N S_i^\al \; , \qquad 
S_\pm \equiv S_x \pm i S_y \; ,
$$
where $\al=x,y,z$ and $S_i^\al$ is an $\al$-component of the spin-1/2
operator. For instance, the population difference (20) can be written 
as the average
$$
s = \frac{2}{N}\; <S_z> \; .
$$
The squeezing factor may be defined [40] as
\be
\label{32}
Q_z \equiv \frac{2\Dlt^2(S_z)}{\sqrt{<S_x>^2+<S_y>^2}} =
\frac{2\Dlt^2(S_z)}{|<S_\pm>|} \; ,
\ee
where $\Dlt^2(S_z)\equiv<S_z^2>-<S_z>^2$. One says that $S_z$ is 
squeezed with respect to $S_\pm$ if $Q_z<1$. Calculating the factor (32), 
we find
\be
\label{33}
Q_z =\sqrt{1-s^2} \; .
\ee
Since $s^2\leq 1$, one has $Q_z\leq 1$. The maximal squeezing is 
achieved when one of the modes is completely populated, i.e. $s=\pm 1$.
Squeezed $S_z$ means that the population difference can be measured with
a better accuracy than the phase difference.

\subsection{Multiparticle entanglement}

The system of interacting atoms possessing internal states may form 
multiparticle entangled states [41,42]. In our case, atoms have not 
internal but collective coherent modes. This makes the principal 
difference of the resonant Bose condensate having collective nonlinear
modes from atoms with internal single-particle states. But the resonant
Bose condensate also forms a multiparticle entangled state. This follows
from the Schr\"odinger representation for the wave function of $N$ atoms
in the resonant condensate,
\be
\label{34}
\Psi_N(\br_1,\br_2,\ldots,\br_N,t) = c_1(t) \prod_{i=1}^N \vp_1(\br_i)
+ c_2(t)\prod_{i=1}^N \vp_2(\br_i) \; ,
\ee
where $\vp_j(\br)$ is a coherent $j$-mode, with $j=1,2$. The function
(34) cannot be factorized into a product of single-particle functions,
provided that $c_1\neq 0$ and $c_2\neq 0$. The maximal entanglement is
reached for $|c_1|=|c_2|=1/\sqrt{2}$, that is, when $s=0$. Since the
coefficients $c_1(t)$ and $c_2(t)$ are defined by the evolution equations 
(12) or (25), from where
\be
\label{35}
|c_1|^2 = \frac{1-s}{2} \; , \qquad |c_2|^2 =\frac{1+s}{2} \; ,
\ee
their values can be manipulated by applying the resonant field and
switching it off at the appropriate moment. In this way, it is possible
to create any degree of entanglement and then one can disentangle the 
created entangled state. The resonant field, thus, acts as a quantum 
disentanglement eraser [43].

\section{Conclusion}

In trapped Bose condensates, the effect of coherent resonance is 
feasible. This is achieved by applying an alternating field with the 
frequency tuned to the transition frequency between two nonlinear modes.
To preserve the resonance condition requires that the number of atoms be
limited by a maximal value. Estimates show that this value is around 
$10^3$ for a typical spherical trap and for alkali atoms. The limiting
number is higher for cigar-shape and disk-shape traps, and can be about
$10^5$ atoms. This number also increases for higher excited nonlinear
modes. Temporal behaviour of fractional populations exhibits dynamic 
critical phenomena occurring on a critical line in the parametric
manifold. The origin of these critical phenomena is the saddle separatrix
crossing by the starting point of a trajectory. Actually, there exists
a whole bunch of critical lines, each of which corresponds to a different
starting point.

As a result of coherent resonance in a system of Bose-condensed atoms, 
a new type of matter is formed, which can be called {\it resonant Bose 
condensate}. This is analogous to a resonant atom, with the principal
difference that the resonant condensate is a coherent multiparticle
system. Collective nature of the latter causes the appearance of
interference patterns, interference current, atomic squeezing, and of
multiparticle coherent entanglement.

We have considered here a single-component Bose gas. Possible extensions 
of the theory could be to binary mixtures of Bose condensates and, 
generally, to multicomponent condensates composed of different atomic 
species. Another possible generalization could be to the systems of cold 
atoms in optical lattices [44] and to radiating atoms interacting with 
electromagnetic field [45]. We think that the effect of coherent 
resonance can find numerous applications, such as creating selected 
modes of atoms lasers, information processing, and quantum computing.

\vskip 5mm

{\bf Acknowledgement}

\vskip 2mm

The work was accomplished in the Research Center for Optics and
Photonics, University of S\~ao Paulo, S\~ao Carlos. Financial support 
from the S\~ao Paulo State Research Foundation is appreciated.

\newpage

\begin{center}

{\large{\bf Figure Captions}}

\end{center}

\vskip 2cm

{\bf Fig. 1}. The population difference $s(t)$ (dashed line) and phase
difference $x(t)$ (solid line) as functions of time, measured in units 
of $\al^{-1}$, for the zero detuning $\ep=0$, initial conditions 
$s_0=-1$, $x_0=0$, and varying amplitude of the resonant field: (a)
$b=0.470$; (b) $b=0.495$; (c) $b=0.501$; (d) $b=0.7$; (e) $b=1$.

\vskip 1cm

{\bf Fig. 2}. Temporal behaviour of the population difference $s(t)$ 
(dashed line) and phase difference $x(t)$ (solid line), under the same
consitions as for Fig. 1, but for the negative parameter $b<0$ taking
the values: (a) $b=-0.40$; (b) $b=-0.49$; (c) $b=-0.6$; (d) $b=-1$.

\vskip 1cm

{\bf Fig. 3}. The population difference $s(t)$ (dashed line) and phase
difference $x(t)$ (solid line) as functions of dimensionless time for 
$\ep=0$ and the initial conditions $s_0=-0.8$, $x_0=0$, with varying 
the parameter $b$ as: (a) $b=-1$; (b) $b=-0.85$; (c) $b=-0.8001$ 
(slightly below the critical point $b_{c1}$); (d) $b=-0.7999$ (slightly 
above the critical point $b_{c1}$); (e) $b=-0.5$; (f) $b=0.18$; (g) 
$b=0.1999$ (slightly below the critical point $b_{c2}$); (h) $b=0.2001$ 
(slightly above the critical point $b_{c2}$); (i) $b=0.5$; (j) $b=1$.  

\vskip 1cm

{\bf Fig. 4}. Temporal behaviour of $s(t)$ (dashed line) and $x(t)$ 
(solid line) for the initial conditions $s_0=-1$, $x_0=0$, fixed $b=-1$,
and a nonzero detuning: (a) $\ep=-0.05$; (b) $\ep=-0.1$.

\vskip 1cm

{\bf Fig. 5}. Phase portrait on the plane $\{ s,x\}$ for a finite 
detuning $\ep=0.1$ and different values of the pumping-field amplitude:
(a) $b=0.51$; (b) $b=0.8$.

\end{document}